\documentclass{frontiersFPHY} 

\usepackage{hyperref}

\def\keyFont{\fontsize{8}{11}\helveticabold }
\def\firstAuthorLast{Fraix-Burnet {et~al.}} 
\def\Authors{Didier Fraix-Burnet\,$^{1*}$, Mauro D'Onofrio\,$^{2}$, Paola Marziani\,$^{3}$}
\def\Address{
$^{1}$UGA / CNRS / IPAG, F-38000 Grenoble, France \\
$^{2}$INAF, Osservatorio Astronomico di Padova, Italy \\
$^{3}$Dipartimento di Fisica \& Astronomia, Universit\`a di Padova, Italy}
\def\corrAuthor{Didier Fraix-Burnet}
\def\corrAddress{Institut de Plan\'etologie et d'Astrophysique de Grenoble (UMR 5274),
BP 53,
F-38041 Grenoble C\'edex 9,
France}
\def\corrEmail{didier.fraix-burnet@obs.ujf-grenoble.fr}

\newcommand{\mnras}[1]{Monthly Notices of the Royal Astronomical Society}
\newcommand{\apj}[1]{The Astrophysical Journal}
\newcommand{\apjl}[1]{The Astrophysical Journal Letters}
\newcommand{\apjs}[1]{The Astrophysical Journal Supplement Series}
\newcommand{\aj}[1]{The Astronmical Journal}
\newcommand{\aap}[1]{Astronomy \& Astrophysics}
\newcommand{\aaps}[1]{Astronomy \& Astrophysics Supplement Series}
\newcommand{\araa}[1]{Annual Review Astrononmy \& Astrophysics}

\begin{document}
\onecolumn
\firstpage{1}

\title[Phylogenetic Analyses of Quasars]{Phylogenetic Analyses of Quasars and Galaxies}
\author[\firstAuthorLast ]{\Authors}
\address{}
\correspondance{}
\extraAuth{}
\topic{}

\maketitle

\begin{abstract} 

\section{}

   Phylogenetic approaches have proven to be useful in astrophysics. We have recently published a Maximum Parsimony (or cladistics) analysis on two samples of 215 and 85 low-z quasars (z $< 0.7$) which offer a satisfactory coverage of the Eigenvector 1-derived main sequence. Cladistics is not only able to group sources radiating at higher Eddington ratios, to separate radio-quiet (RQ) and radio-loud (RL) quasars and properly distinguishes core-dominated and lobe-dominated quasars, but it suggests a black hole mass threshold for powerful radio emission as already proposed elsewhere. An interesting interpretation from this work is that the phylogeny of quasars may be represented by the ontogeny of their central black hole, i.e. the increase of the black hole mass. However these exciting results are based on a small sample of low-z quasars, so that the work must be extended. We are here faced with two difficulties. The first one is the current lack of a larger sample with similar observables. The second one is the prohibitive computation time to perform a cladistic analysis on more that about one thousand objects. We show in this paper an experimental strategy on about 1500 galaxies to get around this difficulty. Even if it not related to the quasar study, it is interesting by itself and opens new pathways to generalize the quasar findings.

\tiny
 \keyFont{ \section{Keywords:} Unsupervised Classification --  Quasars -- Galaxies -- Multivariate Analysis -- Phylogenetic Methods} 
\end{abstract}


\section{Introduction: Astrocladistics}

Astrocladistics\footnote{\url{https://astrocladistics.org}} \citep[][and references therein]{jc1,jc2,FCD06,Fraix-BurnetHouches2016,StatsRef2017,RealmDFB2016} aims at introducing phylogenetic tools in astrophysics.
 
These tools try to establish the relationships between the species by minimizing the total evolutionary cost depicted on a phylogenetic tree. The most general and the simplest to implement technique is Maximum Parsimony, also known as cladistics, and is based on the parameters, and not on distances between the objects.  The trees that result from cladistic analysis should not be interpreted as genealogic trees: here, as the trees do not indicate ancestor or descendant objects, each quasar supposedly represents a species (i.e., a class). In this phylogenetic sense, the trees can be “rooted” according to a parameter that may have an evolutionary meaning. 

The phylogenetic tools are devised to take the evolution of object populations into account. They do not rely on similarities, derived from the computation of distances, but on the fact that diversity is gained through evolution and speciation. For instance, similarity techniques (like most statistical clustering and classification or phenetic tools) tend to find hyperspheres in the parameter space, while phylogenetic tools are able to detect evolutionary paths as can be shown on stellar evolutionary tracks \citep{Fraix-BurnetHouches2016}. Many applications have been published on many kinds of astrophysical objects \citep{FDC09,Fraix2010,Fraix2012,Cardone2013,FD15,Jofre2017,Holt2017}.

Phylogenetic approaches represent the relationships using trees or networks, the first ones being simpler to read. From these evolutionary schemes, it is possible to gather objects into groups that supposedly share the same common ancestor species (monophyletic groups). These groups appear as sub-structures (i.e. bunches of branches) in the tree, their exact number depending on the desired level of details in their physical interpretation.

In this paper, we summarize an exciting cladistic analysis of low-$z$ quasars and illustrate a possible approach to extend such study on much larger samples. 

\section{A cladistic analysis of a low-z quasar sample}

\begin{figure}[t]
\begin{center}
\includegraphics[width=9cm]{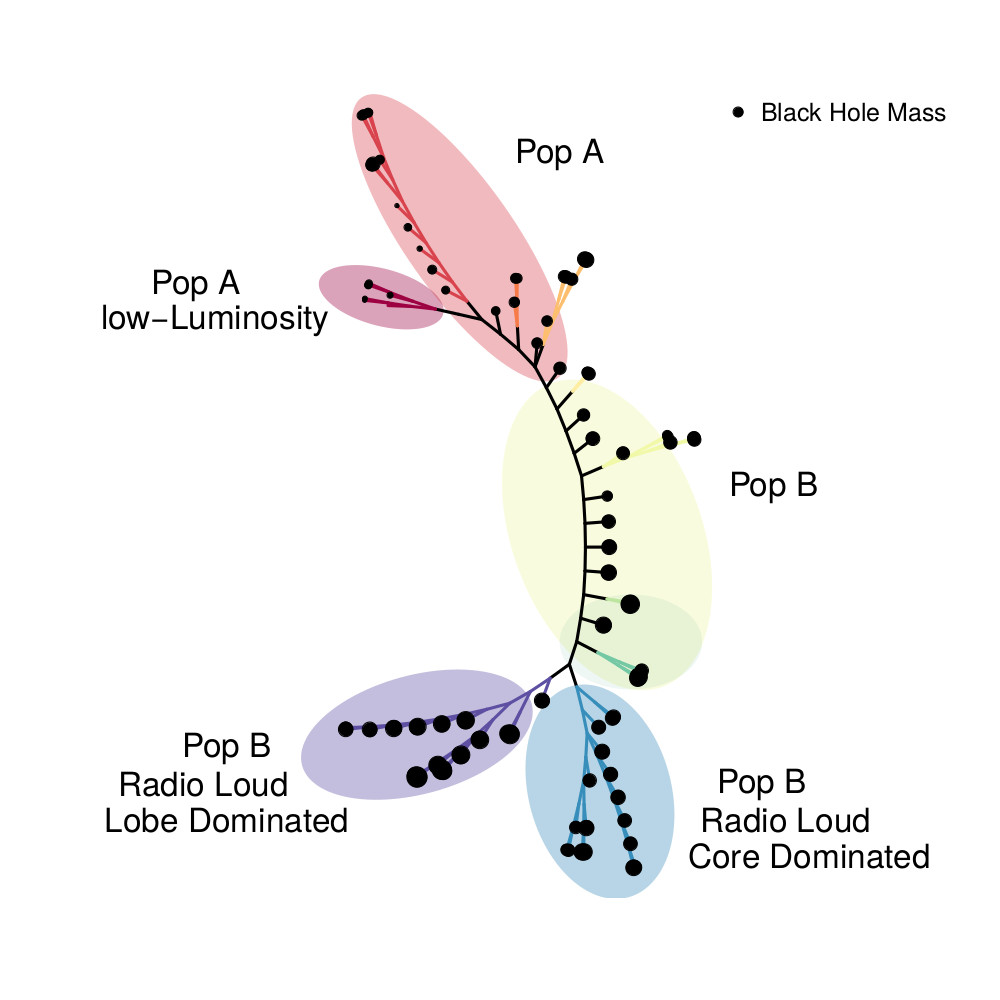}%
\end{center}
 \textbf{\refstepcounter{figure}\label{fig:QSOtree} Figure \arabic{figure}. }{The cladistics tree of 85 quasars. The tree representation is unrooted, but the low black hole masses (MBH) are at the top. We identify ten groups corresponding to bunches of branches and colored as in Fig.~\ref{fig:QSOboxplots}. The colored ellipsoidal regions indicates well known categories of quasars and encompass several of our groups. }
\end{figure}

\begin{figure}[t]
\begin{center}
\includegraphics[width=\linewidth]{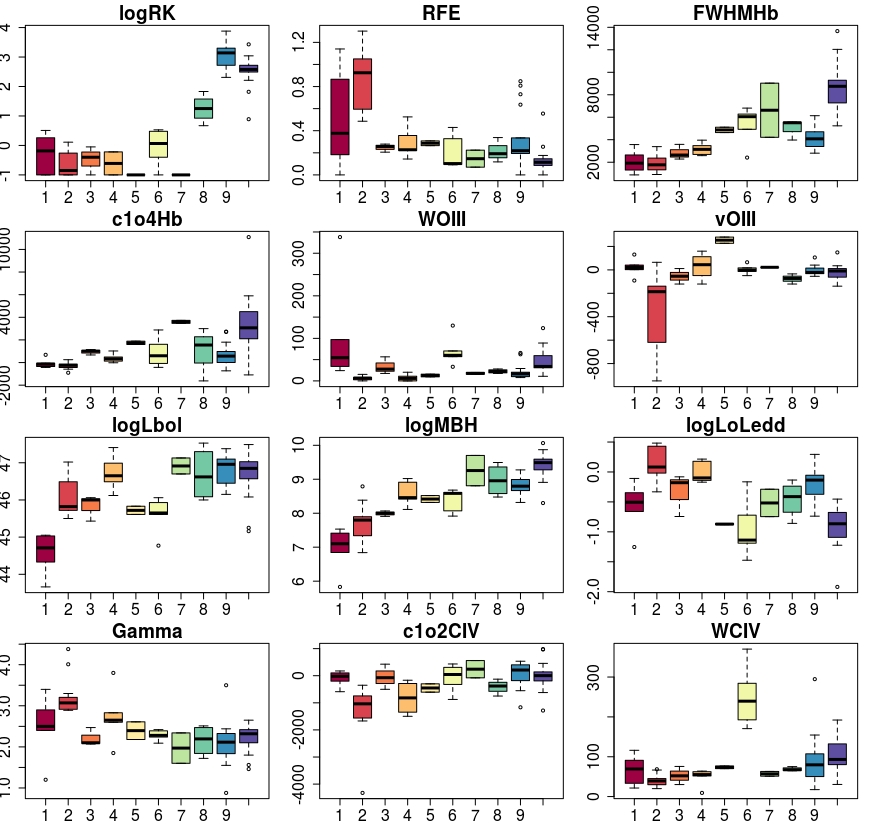}%
\end{center}
 \textbf{\refstepcounter{figure}\label{fig:QSOboxplots} Figure \arabic{figure}. }{Boxplots for the groups in the 85 quasar sample as defined on the tree in Fig.~\ref{fig:QSOtree}. The parameters shown are: the radio loudness parameter (RK), the intensity ratio between Fe$\lambda$4570 and H$\beta$ (RFE), the Full Width at Half-intensity Maximum (FWHMHb) and the line centroid displacement  at quarter maximum (c1o4Hb) of the H$\beta$ line, the equivalent width (WOIII) and the peak shift (vOIII) of the [OIII]$\lambda$5007 line, the bolometric luminosity (Lbol), the Black Hole Mass (MBH), the Eddington ratio L/L$_{edd}$ (LoLedd), the soft X-ray photon index (Gamma), the centroid displacement  at half maximum (c1o2CIV) and  the equivalent width of the CIV$\lambda$1549 line (WCIV).  }
\end{figure}

This analysis is published in \citep{Fraix-Burnet2017}. Two samples of low-redshift (z$\le 0.7$) quasars are used: one with 215 objects presented by \citep{Marziani2003}, and another one made of 85 quasars cross-matched with \citep{Sulentic2007} have measurements of the CIV line. These two samples are modest in size but have good quality measurements of emission lines (H$_{\beta}$, FeII, [OIII], CIV...). For the cladistic analysis, the 215 and 85 object samples have respectively seven and eleven parameters.

With such relatively small samples, the cladistic analysis is relatively easy, and allows for extensive test of its reliability through kinds of bootstrap approaches. The most parsimonious tree in Fig.~\ref{fig:QSOtree} shows the 85 quasars at the leaves (ends of the branches). Bunches of branches that appear to depart from the main trunk are coloured to define groups of quasars that hypothetically may share similar evolutionary histories.

To understand and interpret this tree, it is necessary to look at the properties of the groups, for instance using boxplots (Fig.~\ref{fig:QSOboxplots}). The tree (Fig.~\ref{fig:QSOtree}) is arbitrarily presented with the group having the lowest black hole mass is at the top. The groups on the boxplots are then ordered from the top of the tree to the bottom.

It is striking to note that the black hole mass increases nearly regularly towards the bottom of the tree. Since the black hole mass (MBH) can only grow as a function of quasar evolution and cosmic time, the ontogeny of black holes is represented by their monotonic increase in mass. Considering that MBH provides a sort of “arrow of time” of  nuclear activity, a phylogenetic interpretation of the tree becomes possible if the cladistic tree is rooted on black hole mass.   

Considering other properties, the cladistic tree is thus consistent with the more massive radio-quiet Population B sources (disk dominated, lower Eddingon ratio) at low-z appearing as a more evolved counterpart of  Population A (wind dominated sources, higher Eddington ratio) to which the “local” Narrow-Line Seyfert 1s belong.

The core-dominated  and lobe-dominated Radio Loud (RL) sources are in two distinct groups at the bottom of the tree, indicating they are monophyletic groups. Quite interestingly, these powerful RL sources appear in our low-z sample only above a mass threshold. 

In conclusion, the quasar sample studied in \citep{Fraix-Burnet2017} contains a population of massive quasars which are “more evolved” and a population of less-massive quasars that are radiating at a higher  L/Ledd. While   L/Ledd remains the physical factor governing E1 \citep{Marziani2001,Sulentic2011,Sulentic2015}, high-MBH quasars may have resembled low-MBH quasars in an earlier stage of their evolution.

The cladistic analysis is thus able to recover well-known classes of quasars, but more importantly brings a unique insight on their phylogeny. However, this picture is only valid for the low-z sample studied, and no generalization to the entire quasar population is possible. But the results are sufficiently exciting to justify extensions of this work to other samples. Two directions are foreseen, both requiring higher-z quasar populations, to better depict the quasar evolution. Firstly, it would be interesting to study a sample within a constrained redshift range at another epoch of the Universe to check whether the evolution of the properties of quasars is similar. Secondly, the relationships between quasar populations in a larger redshift range would give a clearer picture of the global evolution of the black hole mass and the different properties like the radio loudness or the disk/wind dichotomy. 

Unfortunately, data are either not existing or of insufficient quality which requires dedicated surveys with large-collecting area telescopes to match the luminosity range of low-z quasars that remain almost unobserved 
at intermediate-to-high redshift \citep{Sulentic2014}. In addition, the cladistic technique is very demanding in computing resources. Basically, all possible trees made of the objects must be built to select the most parsimonious one in terms of evolutionary complexity. There are heuristic tricks to avoid this thorough quest, but still a cladistic analysis is practically not feasible with more than about a thousand objects. 

Another approach is required. Since we do not have a big quasar sample, we present in the following a tentative strategy on a different sample made of galaxies, as an example of potential applications of cladistics to large samples of sources.

\section{A cladistic analysis of low-L ELGs in cluster}

\begin{figure}[ht]
\begin{center}
\includegraphics[width=8cm]{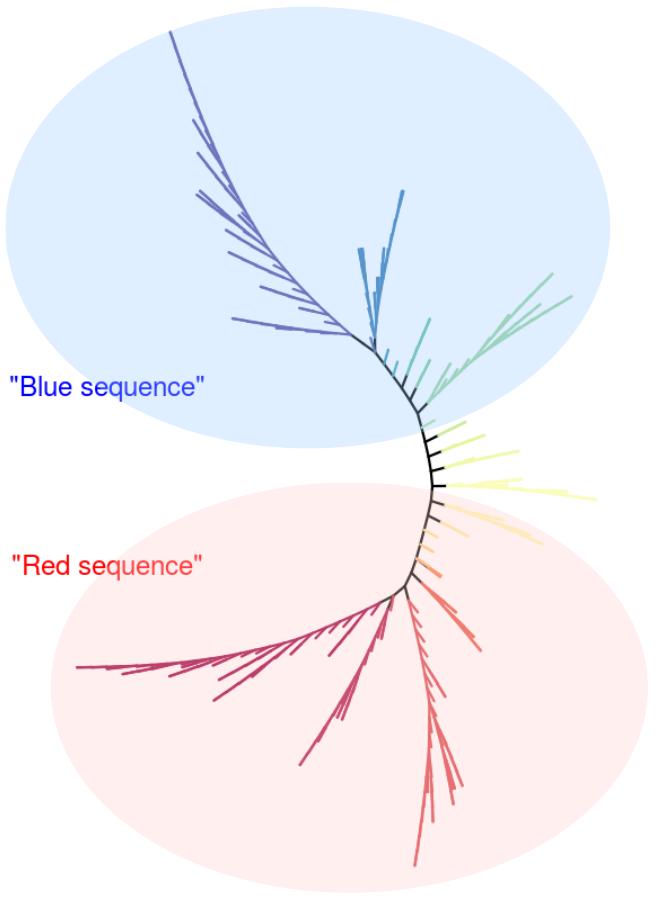}%
\end{center}
 \textbf{\refstepcounter{figure}\label{fig:WINGStree} Figure \arabic{figure}. }{The cladistics tree of 300 pre-clusters of the WINGS sample of 1494 galaxies. The tree representation is unrooted, but the lower masses are at the top. The colors of the branches correspond to the groups and are the same as in Fig.~\ref{fig:WINGSboxplots}.  }
\end{figure}

\begin{figure}[ht]
\begin{center}
\includegraphics[width=\linewidth]{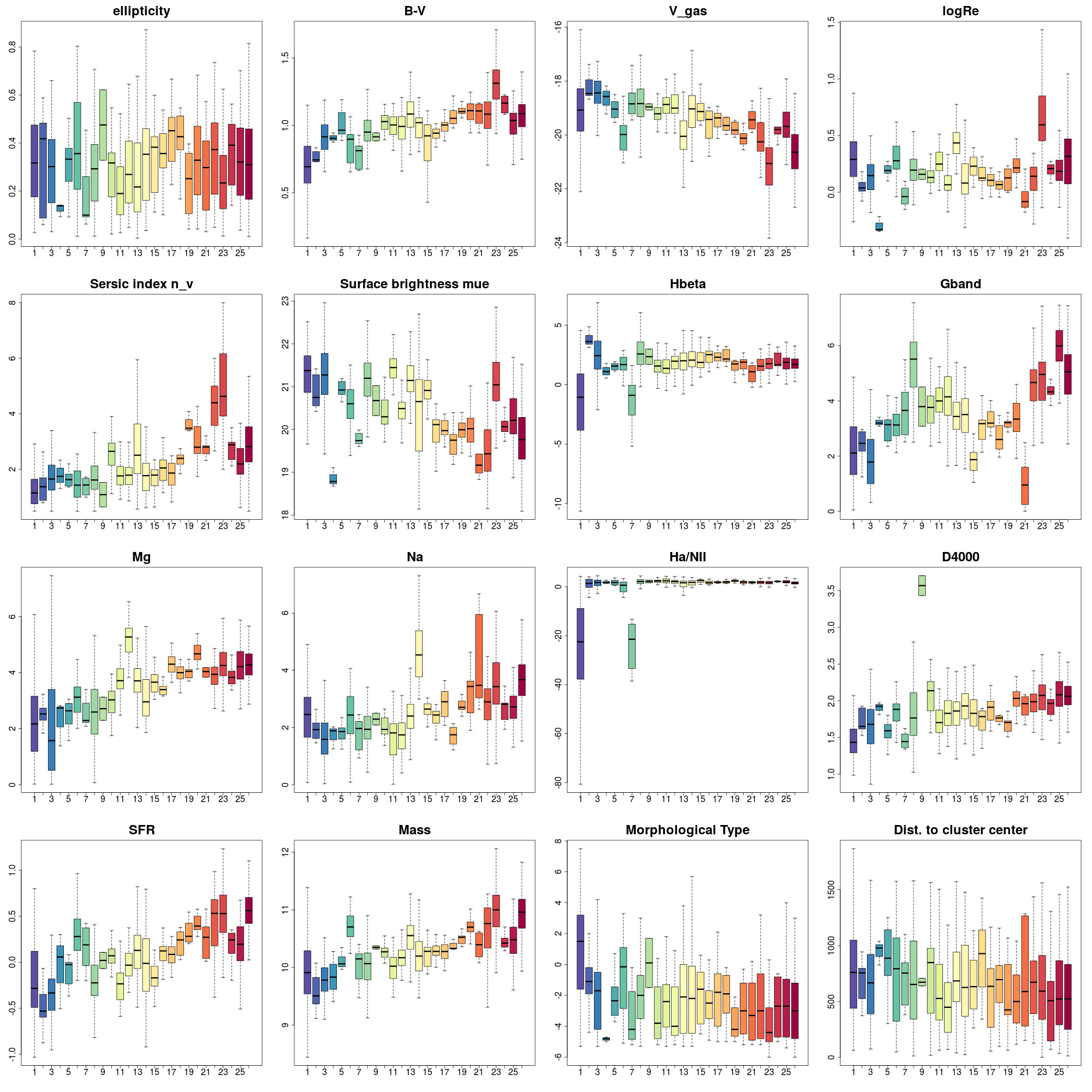}%
\end{center}
 \textbf{\refstepcounter{figure}\label{fig:WINGSboxplots} Figure \arabic{figure}. }{Boxplots for the groups in the WINGS sample as defined on the tree in Fig.~\ref{fig:WINGStree}  }
\end{figure}

\subsection{Sample}

The WINGS survey  \citep{Fasano2006} is an imaging and spectroscopic study of the brightest X-ray clusters at redshift $0.04 <$ z $< 0.07$ selected from the ROSAT all sky survey. The sample for this analysis has 1494 galaxies belonging to several clusters, and eleven parameters have been used for the cladistic analysis itself: B-V, logRe, surface brightness, H$\beta$, D4000, Mass, Sersic index n (measures the degree of curvature of the Sersic profile describing how the intensity of a galaxy varies with distance from its center), H$\alpha$/NII, Gband, Mg and Na. 

\subsection{Pre-clustering}

Phylogenetic methods are intended to find relationships between classes (species) of objects. But there is no multivariate classification of galaxies \citep{Fraix-Burnet2015}. This would be however useful since it is easier and physically more relevant to study different types of objects rather than millions of individuals. This dimension reduction is also necessary in the era of the huge databases brought by current and future telescopes. 

This multivariate classification is one objective of astrocladistics. However, we are limited by the size of the samples to study. There are other phylogenetic techniques that tackle this problem efficiently, but they are based on distances, and most often adapted for the specific evolutionary processes of living organisms and their traits (e.g. \citep{NJ1987,NJ2006}). Some work should be done to assess their applicability to astrophysics. There are also many statistical tools for unsupervised classification (or clustering, \citep{De2013}), but they gather objects according to their similarities, not to their evolutionary relationships. 

We will discuss this big issue with possible solutions in another paper (Fraix-Burnet in prep.), and here  show the results of a first approach we have implemented. 

The idea behind this approach is rather intuitive: we are looking for structures in the parameter space, structures that both gather and relate the objects of our sample. Since we have too many of these objects, we try to reduce the resolution of our data by replacing very close (similar) objects into meta-objects that we call pre-clusters. These pre-clusters take the median properties of their components. In other words, we postulate that there may be some redundancies in our data. Then we can perform the cladistic analysis on these pre-clusters that can subsequently gathered into groups from the tree.

This idea is also mentioned by \citep{Murtagh2014} that recommends to perform a pre-clustering using a hierarchical classification method (that builds a hierarchy of clusters, \citep{Fraix-Burnet2015}) for the k-means analysis (a partitioning method, \citep{kmeans1967,Fraix-Burnet2015}). While for our problem many pre-clustering algorithms could a priori be used, we here choose the hierarchical clustering one. Note that this technique requires a huge amount of CPU time with very large samples.

The number of pre-clusters is arbitrary. Obviously it should not be too low otherwise we probably mix together different kinds of objects. It cannot be too high either because of the limitation of the cladistics analysis. We have found that 300 pre-clusters is here a good choice compromise because the cladistic analaysis takes only a few hours allowing many runs to test this strategy.

\subsection{Results}

The tree (Fig.~\ref{fig:WINGStree}) is obtained with the 300 pre-clusters. Each leave (ending branch) of the tree is thus one pre-cluster. We have gathered these pre-clusters depending on the substructures of the tree, and the groups are represented by different colors. Each group of galaxies thus corresponds either to a single branch or to a bunch of branches on the tree. 

The boxplots (Fig.~\ref{fig:WINGSboxplots}) show the statistics of several parameters for each of the groups. The order of the groups is arbitrary and has been chosen to underline the increase of the mass. On the tree in Fig.~\ref{fig:WINGStree} this parameter increases from top to bottom.

The color progression from blue to red grossly matches the increase in mass of galaxies, as well as other clear trends as visible on the boxplots. Interestingly, the morphological type decreases along the tree downward, and possibly the distance to the cluster center as well even though the in-group scatters are large.

Sometimes, some groups stand out from these trends in some parameters, such as the group 6 for the mass or group 12 for Mg. These groups will be further investigated since they could be either the results of some weird data or, more interestingly, a new peculiar species of galaxies that could lead us to a better understanding of the evolution of galaxies than simply redder colors or larger masses. 

The WINGS sample galaxies belong to X-ray brights clusters, which are rather evolved systems, predominantly close to a state of viral equilibrium. We find that most of the groups have a representant in all the clusters, or conversely all clusters span the entire tree. Despite the low statistics in some of the clusters, this would indicate that the classification scheme depicted on the tree in Fig.~\ref{fig:WINGStree} could be well of general validity for galaxies at low redshift.

We have also analysed a control sample of 497 higher redshift field galaxies. It is impossible to root the tree such that the boxplots show as many monotonic trends as for the cluster sample above, indicating that the field galaxies of our sample may not possess a "common ancestor", that is they could be made of two distinct populations with different origins. Another possibility is that their evolution is more complex, but the sample is probably too small to conclude in this direction.
The fact that the cluster sample of low-redshift galaxies is compatible with a common ancestor can be due to: i) the general influence of clusters on galaxy evolution, ii) time smoothing out somewhat the different origins of these galaxies, iii) a lower diversity by a sort of volume selection effect.

\section{Conclusion}

Categorizing quasars or galaxies is usually made through a handful of properties at most. Multivariate clustering is still rare \citep{Fraix-Burnet2015}, but only phylogenetic tools like cladistics provide relationships that emerge from the data. Here \citep{Fraix-Burnet2017}, the quasar sample is small and relatively well contained in redshift, so probably in diversity. Indeed, the  diversity of the quasar sample (which is exclusively low-$z$, $z \lesssim 0.7$) can be organized along a 1D sequence, the eigenvector-1 main sequence. 

To extend this study to larger samples at higher redshifts, we present a possible strategy to perform the same kind of analysis by performing a pre-clustering using a hierarchical clustering technique, followed by a cladistic analysis on the pre-clusters. This is applied on a galaxy samples due to the current lack of larger samples of quasars with similar parameters as above. 

Even though this diversity is much larger for the WINGS galaxy sample, our proposed strategy successfully establishes a phylogenetic scheme that points to several evolutive properties (like color, mass, metallicity but also D4000 and the Sersic index n) characterizing a level of diversification (or evolution). Some of these evolutive correlations are very probably not causal, unlike the quasar evolution with MBH. 

Some caution is necessary when interpreting the cladograms presented here. One should not conclude that every quasar or every galaxy follows some linear evolution along the tree. There are bunches of branches (sub-structures of the trees) that could suggest some dead ends, or the lack of more ancestral objects. For instance, starting from the low luminosity Pop A quasars (group 1), how to understand the branch of more luminous Pop A quasars (group 2)? Regarding WINGS galaxies, the true ancestors of the objects studied here are at higher redshifts:  where the connection to the presented trees would take place? This is impossible to answer these questions without further pursuing the present work.

\section*{Disclosure/Conflict-of-Interest Statement}

The authors declare that the research was conducted in the absence of any commercial or financial relationships that could be construed as a potential conflict of interest.

\section*{Author Contributions}

The contributions is mainly as follows: DFB performed the cladistic analyses, MDO and PM worked on the WINGS project and gathered the sample. PM also provided the quasar sample. All three authors equally participated to the elaboration of the documents.



%

\bibliographystyle{frontiersinHLTH&FPHY} 
\bibliography{PhyloQuasars}



\end{document}